\documentclass[print]{revtex4}
\usepackage{amsmath,amssymb}
\usepackage{graphicx}
\newcommand{\lvec}[1]{|#1\!\!>}

\draft

\begin{document}

\title{Generation of Two-Flavor Vortex Atom Laser from a Five-State Medium}

\author{Xiong-Jun Liu$^{a,b}$\footnote{Electronic address: phylx@nus.edu.sg},
Hui Jing$^{c}$, Xin Liu$^{a,b}$ and Mo-Lin Ge$^{a,b}$}
\affiliation{a. Theoretical Physics Division, Nankai Institute of
Mathematics,Nankai University, Tianjin 300071, P.R.China\\
b. Liuhui Center for Applied Mathematics, Nankai
University and Tianjin University, Tianjin 300071, P.R.China\\
c. State Key Laboratory of Magnetic Resonance and
Atomic and Molecular Physics,\\
 Wuhan Institute of Physics and Mathematics, CAS, Wuhan 430071, P. R. China}

\begin{abstract}
Two-flavor atom laser in a vortex state is obtained via
electromagnetically induced transparency (EIT) technique in a
five-level $M$ type system by using two probe lights with $\pm
z$-directional orbital angular momentum $\pm l\hbar$,
respectively. Together with the original transfer technique of
quantum states from light to matter waves, the present result can
be extended to generate continuous two-flavor vortex atom laser
with non-classical atoms.

\end{abstract}
\maketitle

\section{introduction}

\indent Since the remarkable observations of Bose-Einstein
Condensation (BEC) in dilute atomic clouds in 1995, there have
been many interests in preparing a continuous atom laser
\cite{atom laser} and exploring its potential applications in,
e.g., gravity measurements through atom interferometry
\cite{interferometer}. Although a sub-quantum-noise atom laser is
expected to be crucial to improve the interferometer
sensitivities, the difficulties for the atomic beam to propagate
over a long distance heavily restrict its actual performance
\cite{noise}. Some time ago Drummond et al. proposed to use
mode-locking technique to stabilize the atom laser based on the
generation of a dark soliton in a ring-shaped condensate
\cite{Drummond}. Other related works are the atomic soliton
formation and its stationary transmission in a traveling optical
laser beam \cite{Zhang} or a waveguide \cite{Leboeuf} for a dense
atomic flow. Also, an optical scheme for generating the soliton
atom laser via electromagnetically induced transparency (EIT)
\cite{1} was proposed in our recent work \cite{sal}.

The generation of a vortex condensate with a Raman adiabatic
passage scheme has been early proposed in the three-level system
\cite{Raman1}. A proposal to create a skyrmion in a Bose-Einstein
Condensate with an $F=1$ hyperfine ground state manifold (three
ground states) using Raman transitions has been made by Marzlin et
al \cite{skyrmion}, and the importance of a tight external trap
for a high efficiency transfer in such a system was analyzed in
detail \cite{Raman2}. In these techniques a sufficiently large
one-photon detuning was needed to avoid the spontaneous emission
from the excited state. Recently, the exchange of orbital momentum
between the light and molecules was also studied by M.Babiker et
al., \cite{orbital}. On the other hand, Juzeliunas and \"{O}hberg
probed the influence of slow light with an orbital momentum
angular on the mechanical properties of three-level atomic
Degenerate Fermi Gases (DFG) and pointed out the existence of an
effective magnetic field and its analogy de Haas-van Alphen effect
in the neutral DFG system \cite{ohberg}. For the three-level case,
to obtain the stable vortex state, one usually requires that the
ratio of Rabi-frequencies between probe and control fields should
be (approximately) independent of space \cite{Raman1,Raman2}. In
this article, we proceed to generate two-flavor atom laser in
vortex states by extending the Juzeliunas-Ohberg model to a
five-level $M$ type atomic system. The equations of motion for the
output atomic beams can be described as that of an effective
two-flavor (oppositely charged) Bose condensate in an effective
magnetic field, and the ratios of Rabi-frequencies in this
technique are not needed to be constant.

The present technique was based on the physical mechanism of
Electromagnetically Induced Transparency (EIT) \cite{1}, which has
attracted much attention in both experimental and theoretical
aspects in recent years \cite{2,3,4}, especially after
Fleischhauer and Lukin formulated their elegant dark-states
polaritons (DSPs) theory \cite{5} and thereby the rapid
developments of quantum memory technique, i.e., transferring the
quantum states of photon wave-packet to collective Raman
excitations in a loss-free and reversible manner. By extending the
transfer technique to matter waves, Fleischhauer and Gong proposed
a wonderful scheme to make a continuous atomic beam with
non-classical or entangled states \cite{6}, which was later
confirmed for double-$\Lambda$ four-level atomic medium
\cite{liu}, even to a soliton atom laser \cite{sal}.

In the following we propose a technique to generate two-flavor
atom lasers in vortex states via induced Raman transitions in a
five-level $M$ type system which interacts with two probe lights
which have $\pm z$-directional orbital angular momentum $\pm
l\hbar$, respectively, and two external control fields. If the
ratio between the Rabi frequencies of the first pair of probe and
control fields equals that of the second pair, a vortex state of
two-flavor atom laser will be obtained. Together with the original
Fleischhauer-Gong scheme \cite{6}, the present result can be
extended to generate continuous two-flavor vortex atom laser with
non-classical atoms. Besides the applications to quantum
information, the present system with effective $\pm$ charge may
have very interesting applications to spintronics
\cite{spincurrent}.

\section{model}
The system we consider is shown in Fig.1(a) \cite{five}. The
condensate atoms with internal five-level $M$ type configuration
interact with four laser beams: Two strong control lasers
respectively drive the transition $|2\rangle\rightarrow|4\rangle$
with Rabi frequency $\Omega_{c_1}=\Omega_{c_1}^{(0)}\exp(i\bold
k_{c_1}\cdot\bold r)$ and $|3\rangle\rightarrow|5\rangle$ with
$\Omega_{c_2}=\Omega_{c_2}^{(0)}\exp(i\bold k_{c_2}\cdot\bold r)$,
where $\Omega_{c_j}^{(0)} (j=1,2)$ are the slowly varying
amplitudes and $k_{c_j}$ the wave-vectors. The two probe fields
coupling the transitions $|1\rangle\rightarrow|4\rangle$ and
$|1\rangle\rightarrow|5\rangle$, respectively, are characterized
by the wave-vectors $\bold k_{p_j}=k_{p_j}\hat{\bold z} (j=1,2)$
and the Rabi frequency
$\Omega_{p_1}=\Omega_{p_1}^{(0)}e^{i(l\phi+\bold k_{p_1}\cdot\bold
r)}$ and $\Omega_{p_2}=\Omega_{p_2}^{(0)}e^{i(l'\phi+\bold
k_{p_1}\cdot\bold r)}$ where $\Omega_{p_j}^{(0)}$ are also the
slowly varying amplitudes. Here $l$ and $l'$ indicate that the
probe fields are assumed to respectively have orbital angular
momentums $\hbar l$ and $\hbar l'$ along the $z$ direction. It is
convenient to introduce the time-slowly-varying amplitudes
$\Psi_2=\Phi_2e^{-i(\omega_{p_1} -\omega_{c_1})t}$,
$\Psi_4=\Phi_4e^{-i\omega_{p_1}t}$,
$\Psi_3=\Phi_3e^{-i(\omega_{p_2} -\omega_{c_2})t}$, and
$\Psi_5=\Phi_5e^{-i\omega_{p_2}t}$. Hence the equations of motion
for the matter fields are given by
\begin{eqnarray}\label{eqn:1}
i\hbar\frac{\partial\Phi_1}{\partial
t}=-\frac{\hbar^2}{2m}\nabla^2\Phi_1+V_1(\bold
r)\Phi_1+[U_{11}|\Phi_1|^2+U_{12}|\Phi_2|^2
+U_{13}|\Phi_3|^2]\Phi_1+\hbar
\Omega_{p_1}^*\Psi_4+\hbar\Omega_{p_2}^*\Phi_5,
\end{eqnarray}
\begin{eqnarray}\label{eqn:2}
i\hbar\frac{\partial\Phi_2}{\partial
t}=(\epsilon_{12}-\frac{\hbar^2}{2m}\nabla^2)\Phi_2+V_2(\bold
r)\Phi_2+[U_{21}|\Phi_1|^2+U_{22}|\Phi_2|^2+U_{32}|\Phi_3|^2]\Phi_1
+\hbar \Omega_{c_1}^*\Phi_4,
\end{eqnarray}
\begin{eqnarray}\label{eqn:3}
i\hbar\frac{\partial\Phi_3}{\partial
t}=(\epsilon_{13}-\frac{\hbar^2}{2m}\nabla^2)\Phi_3+V_3(\bold
r)\Phi_3+[U_{31}|\Phi_1|^2+U_{32}|\Phi_2|^2+U_{33}|\Phi_3|^2]\Phi_1
+\hbar\Omega_{c_2}^*\Phi_5,
\end{eqnarray}
\begin{eqnarray}\label{eqn:4}
i\hbar\frac{\partial\Phi_4}{\partial
t}=(\epsilon_{14}-\frac{\hbar^2}{2m}\nabla^2)\Phi_4+V_4(\bold
r)\Phi_4+\hbar\Omega_{c_1}\Phi_2+\hbar\Omega_{p_1}\Phi_1,
\end{eqnarray}
\begin{eqnarray}\label{eqn:5}
i\hbar\frac{\partial\Phi_5}{\partial
t}=(\epsilon_{15}-\frac{\hbar^2}{2m}\nabla^2)\Phi_5+V_5(\bold
r)\Phi_5+\hbar\Omega_{c_2}\Phi_3+\hbar\Omega_{p_2}\Phi_1.
\end{eqnarray}
where $V_i(\bold r) (i=1,2,3,4,5)$ are the external potentials,
the scattering length $a_{ij}$ characterizes the atom-atom
interactions via $U_{ij}=4\pi\hbar^2a_{ij}/m$ of which for
simplicity we assume the scattering length $a_{ij}=a_0$ is
constant. $\epsilon_{14}=\hbar(\omega_{41}-\omega_{p_1})$ and
$\epsilon_{15}=\hbar(\omega_{51}-\omega_{p_2})$ are energies of
single-photon detunings, while
$\epsilon_{12}=\hbar(\omega_{21}-\omega_{p_1}-\omega_{s_1})$ and
$\epsilon_{13}=\hbar(\omega_{31}-\omega_{p_2}-\omega_{s_2})$ are
energies of the two-photon detunings. Since almost no atoms occupy
the excited state $\lvec{3}$ in the dark-state condition that was
fulfilled in EIT technique, the collisions between two excited
states and lower states can safely be neglected.

Assuming that the two-photon detunings $\epsilon_{12}$ and
$\epsilon_{13}$ are sufficiently small and the strengths of the
probe fields are much smaller than that of the control fields, one
arrives at the adiabatic condition relating $\Phi_2$ and $\Phi_3$
to $\Phi_1$ \cite{ohberg}. Hence, for the dark-state condition
\cite{five,dark-state}, from eq. (\ref{eqn:4}) and (\ref{eqn:5})
we have in the zero order
\begin{eqnarray}\label{eqn:dark-state}
\Phi_2=-\xi_1\Phi_1, \ \ \Phi_3=-\xi_2\Phi_1,
\end{eqnarray}
where the ratio coefficients
$\xi_1==-\frac{\Omega_{p_1}}{\Omega_{c_1}}$ and
$\xi_2=-\frac{\Omega_{p_2}}{\Omega_{c_2}}$. With above relations
one can easily derive the equations of motion for $\Phi_1, \Phi_2$
and $\Phi_3$ \cite{ohberg}. For example, substituting the above
formula into the equations (\ref{eqn:2}) and (\ref{eqn:3}),
respectively, yields
\begin{eqnarray}\label{eqn:rel1}
\Phi_4&=&-(\hbar\Omega_{c_1}^*)^{-1}\bigr\{[i\hbar\frac{\partial}{\partial
t}-(\epsilon_{12}-\frac{\hbar^2}{2m}\nabla^2)+V_2(\bold
r)]\xi_1^{-1}\Phi_1-\nonumber\\
&&-(U_{21}|\Phi_1|^2+U_{22}|\xi_1|^{-2}|\Phi_1|^2
+U_{32}|\xi_2|^{-2}|\Phi_1|^2) \bigr\}\Phi_1 ,
\end{eqnarray}
\begin{eqnarray}\label{eqn:rel2}
\Phi_5&=&-(\hbar\Omega_{c_2}^*)^{-1}\bigr\{[i\hbar\frac{\partial}{\partial
t}-(\epsilon_{13}-\frac{\hbar^2}{2m}\nabla^2)+V_3(\bold
r)]\xi_2^{-1}\Phi_1+\nonumber\\
&&+(U_{31}|\Phi_1|^2+U_{32}|\xi_1|^{-2}|\Phi_1|^2
+U_{33}|\xi_2|^{-2}|\Phi_1|^2)\bigr\}\Phi_1 ,
\end{eqnarray}
Together with the above two relations, from the eq.(\ref{eqn:1})
we can obtain the equation of motion for the field $\Phi_1$.
Similarly, from the eq. (\ref{eqn:1}) and eq. (\ref{eqn:3}) (or
eq.(\ref{eqn:2})) one can derive the relation between field
$\Phi_4$ ($\Phi_5$) and $\Phi_2$ (or $\Phi_3$), and then
substitute it to the eq. (\ref{eqn:2}) (eq.(\ref{eqn:3})) we can
obtain the equation of motion for the field $\Phi_2$ (or
$\Phi_3$). For this we finally find the equations of motion for
the matter fields $\Phi_1$, $\Phi_2$ and $\Phi_3$ read
\begin{eqnarray}\label{eqn:6}
i\hbar\frac{\partial}{\partial
t}\Phi_{\alpha}=\frac{1}{2m}[i\hbar\nabla+\vec{A}_{\alpha}]^2\Phi_{\alpha}
+V_{\alpha{eff}}\Phi_{\alpha}+U|\Phi_1|^2\Phi_{\alpha}, \ \
(\alpha=1,2,3).
\end{eqnarray}
where we have ignored nonlinear terms involving $\Phi_2$ or
$\Phi_3$ for small $|\xi_j|$, the nonlinear interaction strength
$U=4\pi\hbar^2a_0/m$ and the effective vectors
\begin{eqnarray}\label{eqn:7}
\vec{A}_1=\hbar \
\Xi_1^{-1}(\xi_1^*\nabla\xi_1+\xi_2^*\nabla\xi_2),
\end{eqnarray}
\begin{eqnarray}\label{eqn:8}
\vec{A}_2=\hbar \ \Xi_2^{-1}\bigr(1/\xi_2^*\nabla(1/\xi_2)
+\xi_1^*/\xi_2^*\nabla(\xi_1/\xi_2)\bigr),
\end{eqnarray}
\begin{eqnarray}\label{eqn:9}
\vec{A}_3=\hbar \ \Xi_3^{-1}\bigr(1/\xi_1^*\nabla(1/\xi_1)
+\xi_2^*/\xi_1^*\nabla(\xi_2/\xi_1)\bigr),
\end{eqnarray}
where $\Xi_1=1+|\xi_1|^2+|\xi_2|^2,
\Xi_2=1+1/|\xi_2|^2+|\xi_1|^2/|\xi_2|^2$ and
$\Xi_3=1+1/|\xi_1|^2+|\xi_2|^2/|\xi_1|^2$, and the effective trap
potentials $V_{1eff}=i\hbar \
\Xi_1^{-1}[V_1+|\xi_1|^2V_2+|\xi_2|^2V_3
+(2m\Xi_1)^{-1}|\vec{A_1}|^2], V_{2eff}=i\hbar \
\Xi_2^{-1}\bigr[V_2+(\epsilon_{21}+V_1)/|\xi_2|^2-V_3|\xi_1|^2/|\xi_2|^2+(2m\Xi_2)^{-1}|\vec{A_2}|^2\bigr]
$ and $V_{3eff}=i\hbar \
\Xi_3^{-1}\bigr[V_3+(\epsilon_{31}+V_1)/|\xi_1|^2-V_2|\xi_2|^2/|\xi_1|^2+(2m\Xi_3)^{-1}|\vec{A_3}|^2\bigr]$.
One can find that the effective vectors $\vec{A}_{\alpha}
(\alpha=1,2,3)$ is generally non-Hermitian. The Hermitian
contribution is due to the changes in the phase of $\xi_{1,2}$.
and the non-Hermitian part results in the changes of the amplitude
of the bosonic fields $\Phi_{\alpha}$. Noting that the
dimensionless function
$\xi_j=e^{iR_j}\Omega_{p_j}^{(0)}/\Omega_{c_j}^{(0)} (j=1,2)$
where the phases $R_1=(\bold k_{p_1}-\bold k_{c_1})\cdot \bold
r+l\phi$ and $R_2=(\bold k_{p_2}-\bold k_{c_2})\cdot \bold
r+l'\phi$, and under the condition
$\bigr|\nabla|\xi_j|^2\bigr|\ll\bigr||\xi_j|^2\nabla R_j\bigr|$
\cite{ohberg}, one can neglect the non-Hermitian part. The
effective vectors then yield
\begin{eqnarray}\label{eqn:7'}
\vec{A}_1=\hbar \ \Xi_1^{-1}(|\xi_1|^2\nabla R_1+|\xi_2|^2\nabla
R_2),
\end{eqnarray}
\begin{eqnarray}\label{eqn:8'}
\vec{A}_2=-\hbar \ \Xi_2^{-1}\bigr(|1/\xi_2|^2\nabla R_2
-|\xi_1|^2/|\xi_2|^2\nabla(R_1-R_2)\bigr),
\end{eqnarray}
\begin{eqnarray}\label{eqn:9'}
\vec{A}_3=\hbar \ \Xi_3^{-1}\bigr(|1/\xi_1|^2\nabla R_1
-|\xi_2|^2/|\xi_1|^2\nabla(R_2-R_1)\bigr).
\end{eqnarray}

When $|\xi_1|^2=|\xi_2|^2$, i.e., the ratio between the Rabi
frequencies $\Omega_{p_1}^{(0)}$ and $\Omega_{c_1}^{(0)}$ equals
that between $\Omega_{p_2}^{(0)}$ and $\Omega_{c_2}^{(0)}$
(However, one has to keep in mind that the ratio function itself
is never restricted), and the photons of two input probe fields
have opposite orbital angular momentum, i.e. $l=-l'$, one can
easily obtain $\vec{A}_1=0$ and
$\vec{A}_2=-\vec{A}_3=\vec{A}=-\hbar l\nabla\phi$. The equations
of (\ref{eqn:6}) can then be rewritten as
\begin{eqnarray}\label{eqn:13}
i\hbar\frac{\partial}{\partial t}{\Phi_2 \choose
\Phi_3}=\frac{1}{2m}\bigr[i\hbar\nabla+q\sigma_3\vec{A}]^2{\Phi_2
\choose \Phi_3}+{\left[ \begin{array}{lc} V_{2eff} & 0\\
0 & V_{3eff} \end{array} \right]}
 {\Phi_2 \choose
\Phi_3}+U\rho(\bold r){\Phi_2 \choose \Phi_3},
\end{eqnarray}
where $\sigma_3$ is the pauli matrix, $q=+1$, and
\begin{eqnarray}\label{eqn:14}
i\hbar\frac{\partial}{\partial
t}\Phi_{1}=-\frac{1}{2m}\hbar\nabla^2\Phi_{1}+V_{1eff}(\bold
r)\Phi_{1}.
\end{eqnarray}

The above equations can easily be understood that the three
categories of condensates with effective electric charges $q_1=0$
and $q_2=-q_3=q=+1$, respectively, interact with an effective
external magnetic field $\bold B_{eff}=\nabla\times\vec{A}$. In
other words, by using two probe lights that respectively have
orbital angular momentum $\pm l\hbar$ in the $\pm z$ direction, we
obtain an effective two-flavor (oppositely charged) condensate
which attracts special attention in recent years \cite{faddeev}.
For present purpose, we consider the case $|\xi_j|^2\ll1$, and
have neglected the depletion of atoms in the trapped state
$|1\rangle$, and linearized the equation (\ref{eqn:13}) by using
\cite{vortex} $|\Phi_1|\approx\sqrt{\rho(\bold r)}$ with
$\rho(\bold r)$ the total density of the condensate. As it is
well-known, the ideal configuration for an atom laser can be
described as: Using a confining magnetic potential to act as a
cavity for the laser mode, which is occupied by a Bose-Einstein
condensate, radio-frequency or optical Raman transitions are then
used to coherently transferred atoms into untrapped hyperfine
states, which can propagate freely away from the remaining trapped
atoms \cite{laser}. For this we may choose $V_2(\bold r)$ and
$V_3(\bold r)$ such that $V_{2eff}(\bold r)=0$ and $V_{3eff}(\bold
r)=0$. Together with the eq. (\ref{eqn:dark-state}), one can find
the solution of eq. (\ref{eqn:13}) in the following form
\begin{eqnarray}\label{eqn:15}
\Phi_2(\bold
r,t)=-\frac{\Omega_{p_1}^{(0)}}{\Omega_{c_1}^{(0)}}\sqrt{\rho(\bold
r)}\exp{(iS_2(\bold r,t))}, \ \ \Phi_3(\bold
r,t)=-\frac{\Omega_{p_2}^{(0)}}{\Omega_{c_2}^{(0)}}\sqrt{\rho(\bold
r)}\exp{(iS_3(\bold r,t))},
\end{eqnarray}
where the phases $S_2(\bold r,t)=\frac{q_2}{\hbar}\int^{\bold
r}_0\vec{A}\cdot d\bold r'+S_{02}(\bold r,t)$ and $S_3(\bold
r,t)=\frac{q_3}{\hbar}\int^{\bold r}_0\vec{A}\cdot d\bold
r'+S_{03}(\bold r,t)$ with $S_{02}(\bold r,t)=\bold{\bar
k}_2\cdot\bold r-\int^t_0{dt'\bigr(V_{2eff}(\bold r+\bold{\bar
K}_2(t'-t))+U\rho(\bold r+\bold{\bar K}_2(t'-t))\bigr)}$ and
$S_{03}(\bold r,t)=\bold{\bar k}_3\cdot\bold
r-\int^t_0{dt'\bigr(V_{3eff}(\bold r+\bold{\bar
K}_3(t'-t))+U\rho(\bold r+\bold{\bar K}_3(t'-t))\bigr)}$.
$\bold{\bar K}_2=\hbar\bold{\bar k}_2/m=\hbar(\bold k_{p_1}-\bold
k_{c_1})/m$ and $\bold{\bar K}_3=\hbar\bold{\bar
k}_3/m=\hbar(\bold k_{p_2}-\bold k_{c_2})/m$ are the corresponding
recoil velocities. The velocity spread of the out-coupled matter
fields can be obtained by $\vec{v}_j(\bold r)=\hbar\nabla_{\bold
r}S_j(\bold r,t)/m \ (j=2,3)$. It can be verified that the loop
integration of the velocity yields
\begin{eqnarray}\label{eqn:16}
\oint_{z=0\in C_j}\vec{v}_j(\bold r')\cdot d\bold r'=\pm2\pi\hbar
l/m, \ \ (j=2,3),
\end{eqnarray}
where the sign of right-hand side of above equation takes $+$ (for
$j=2$) or $-$ (for $j=3$). $z=0\in C_j$ means that the integration
path $C_j$ encircles the $z$-axis. The above results indicate that
the orbital angular momentum of the input probe fields can be
transferred into the out-coupled two-flavor atom lasers in
corresponding vortex states (fig.1(b)). Practically, since the
small ratio between the Rabi frequency of probe field and that of
control field has a spatially distribution, strictly one can not
obtain a vortex atom laser in usual three-level EIT technique
\cite{ohberg}. Here, with EIT technique in a five-level $M$ type
system, the two-flavor atom lasers in a vortex state can be
generated when the ratio between the Rabi frequencies of the first
pair of probe and control fields ($\Omega_{p_1}^{(0)}$ and
$\Omega_{c_1}^{(0)}$) equals that between the second pair
($\Omega_{p_2}^{(0)}$ and $\Omega_{c_2}^{(0)}$), while the
spatially-dependent character of the ratio function itself is
never restricted. The present system with effective $\pm$ charge
may have very interesting applications. For example, if the
different internal states represent different spin states of the
atoms, and since the vortex states with different effective
charges can be controlled independently by using external
potentials, we can generate a spin current by controlling the
atoms with different vortex states to move to opposite directions.
This will be very useful for spintronics \cite{spincurrent}, which
will be studied specifically in our next publication.

\section{discussion on nonclassical case}
It is noteworthy that the above result can be extended to generate
stationary continuous two-flavor vortex atom laser with
non-classical atoms using the Fleischhauer-Gong technique
\cite{6,liu}. For a brief discussion we consider a beam of
five-level $M$ type atoms moving in $+z$ direction interacts with
two spatially varying control Stokes fields and two quantized
probe fields with $\pm z$-directional orbital angular momentum
$\pm l\hbar$, respectively. The control Stokes fields are taken to
be much stronger than the probe ones. The Rabi-frequencies of the
Stokes fields can be described by $\Omega_j(\bold
r,t)=\Omega_{0j}(\bold r)e^{-i\omega_{s_j}(t-z/c_j)}$ with
$\Omega_{0j}~(j=1,2)$ being taken as real and $c_j$ denoting the
phase velocities projected onto the $z$ axis, the two quantized
probe fields are characterized by the dimensionless positive
frequency components $E^{(+)}_j(\bold r,t)={\cal E}_j(\bold
r,t)e^{-i\omega_{pj}(t-z/c)+il_j\phi} \ (l_1=-l_2=l)$, and the
ratio of the absolute value of Rabi frequencies between the first
pair of probe and Stokes fields equals that between second pair,
i.e. $g_1^2\langle{\cal E}_1^{\dag}{\cal E}_1\rangle/\Omega_{01}^2
=g_2^2\langle{\cal E}_2^{\dag}{\cal E}_2\rangle/\Omega_{02}^2$,
where $g_j$ is the coupling constant \cite{1}. Similar to the
former derivation in this paper, we may introduce the
slowly-varying amplitudes $\Psi_1=\Phi_1(\bold
r,t)e^{i(k_0z-\omega_0t)}$, $\Psi_2=\Phi_2(\bold
r,t)e^{i[(k_0+k_{p_1}-k_{c_1})z-(\omega_0+\omega_{p_1}-\omega_{c_1})t]}$
$\Psi_3=\Phi_3(\bold
r,t)e^{i[k_0+k_{p_2}-k_{c_2})z-(\omega_0+\omega_{p_2}-\omega_{c_2})]}$,
$\Psi_4=\Phi_4(\bold
r,t)e^{-i(k_0+k_{p_1})z-(\omega_0+\omega_{p_1})t}$ and
$\Psi_5=\Phi_5(\bold
r,t)e^{-i(k_0+k_{p_2})z-(\omega_0+\omega_{p_2})t}$, where
$\hbar\omega_0=\hbar^2k^2_0/2m$ is the corresponding kinetic
energy in the mean velocity, $k_{p_j} $ and $k_{s_j}$ $(j=1,2)$
are respectively the vector projections of the two probe and
Stokes fields to the $z$ axis. The atoms have a narrow velocity
distribution around $v_0=\hbar{k_0}/2m$ with
$k_0\gg|k_{p_j}-k_{s_j}| (j=1,2)$ to minimize the effect of
Doppler broadening \cite{liu}. All fields are assumed to be in
resonance for the central velocity class.

We consider a stationary input of atoms in state $|1\rangle$,
i.e., at the entrance region $|\Phi_1(\bold
r,t)|_{z=0}=(\rho(\bold r))^{1/2}|_{z=0}$. If the Rabi-frequencies
$\Omega_{01}$ and $\Omega_{02}$ are sufficiently slowly,
monotonically decreasing function of $z$ but dependent on $x$ and
$y$ and approach zero at the entrance $z=L$ and all the effective
trap potentials are assumed to be zero \cite{6}, together with the
former results, the output matter waves can then be obtained in
the vortex state
\begin{eqnarray}\label{eqn:16}
\Phi_{2,3}(\bold r,t)|_{z=L}=-\sqrt{\frac{c}{v_0}}{\cal
E}_{1,2}\bigr(\bold
r,t-\tau_{2,3}(L)\bigr)|_{z=0}\exp{\bigr[iS_{2,3}(\bold
r,t)\bigr]},
\end{eqnarray}
where $\tau_j(L)=\int^L_0dz'/V^{(j)}_g(z')$ and the group velocity
$V^{(j)}_{g}=c(1+\frac{g^2_jn}{\Omega_{0j}^2}\frac{v_0}{c})
/(1+\frac{g^2_jn}{\Omega_{0j}^2}) \ (j=2,3)$ with
$n=\int{\rho(\bold r)dxdy}$ is the total linear-density of the
condensate \cite{6,liu}. The factor $\sqrt{c/v_0}$ accounts for
the fact that the input light pulses propagate with velocity $c$
and the output matter field propagates with $v_0$. In fact, the
Raman process of present five-level system can be considered as
two separate three-level Raman process, i.e. the process of atom
field $\Phi_1$ to $\Phi_2$ using the first pair probe and control
fields and the process of atom field $\Phi_1$ to $\Phi_3$ using
the second pair probe and control fields. Then, the quantum states
of the quantized probe field ${\cal E}_1$ can be transferred to
the output atom field $\Phi_2$ and the quantum states of probe
field ${\cal E}_2$ can be transferred to the output atom field
$\Phi_3$, respectively. Particularly, when the entangled probe
lights are used, we can obtain a pair of entangled vortex atom
lasers. On other hand, in a certain extent the vortex states may
be helpful to overcome the decoherence effect and other
difficulties that restrict the atomic beam to propagate over a
long distance. For this our result can be useful for quantum
information processing based on the non-classical vortex atom
lasers.

\section{conclusions}
In conclusion we obtain a two-flavor atom laser in a vortex state
via electromagnetically induced transparency (EIT) technique in a
five-level $M$ type system by using two probe lights with $\pm
z$-directional orbital angular momentum $\pm l\hbar$,
respectively. The key point of present technique is that if the
the ratio between the Rabi frequencies of the first pair of probe
and control fields equals that of the second pair, the equations
of motion for the output atomic beams can be described as that of
an effective two-flavor (oppositely charged) Bose condensate in an
effective magnetic field, and then a vortex state of two-flavor
atom laser will be obtained. Together with the original
Fleischhauer-Gong scheme, the present result can be extended to
generate continuous two-flavor vortex atom laser with
non-classical atoms. Finally, it has other very interesting
applications for the superposition state of different circulating
vortex states labelled by the distinctive internal states. For
example, if the different internal state represents different spin
state of the atoms, and since the different vortex state can be
controlled independently by using external potentials, we can
generate a spin current by controlling the atoms with different
vortex states to move to opposite directions. This will be very
useful for spintronics, which will be studied specifically in our
next work.
\\

We thank Lei-Na Zhu and Zheng-Xin Liu for valuable discussions.
This work is supported by NUS academic research Grant No. WBS:
R-144-000-071-305, and by NSF of China under grants No.10275036
and No.10304020.

%\section{Results}
%\section{Conclusions}
%\indent
%ÕýÎÄ

%%%%%%%%%%%%%%%%%%%%%%%%%%%%%%%%%%%%%%%%%%%%%

%%%%%%%%%%%%%%%%%%%%%%%%%%%%%%%%%%%%%%%%%%%%%
\noindent\\  \\

\begin{figure}[ht]
\includegraphics[width=0.6\columnwidth]{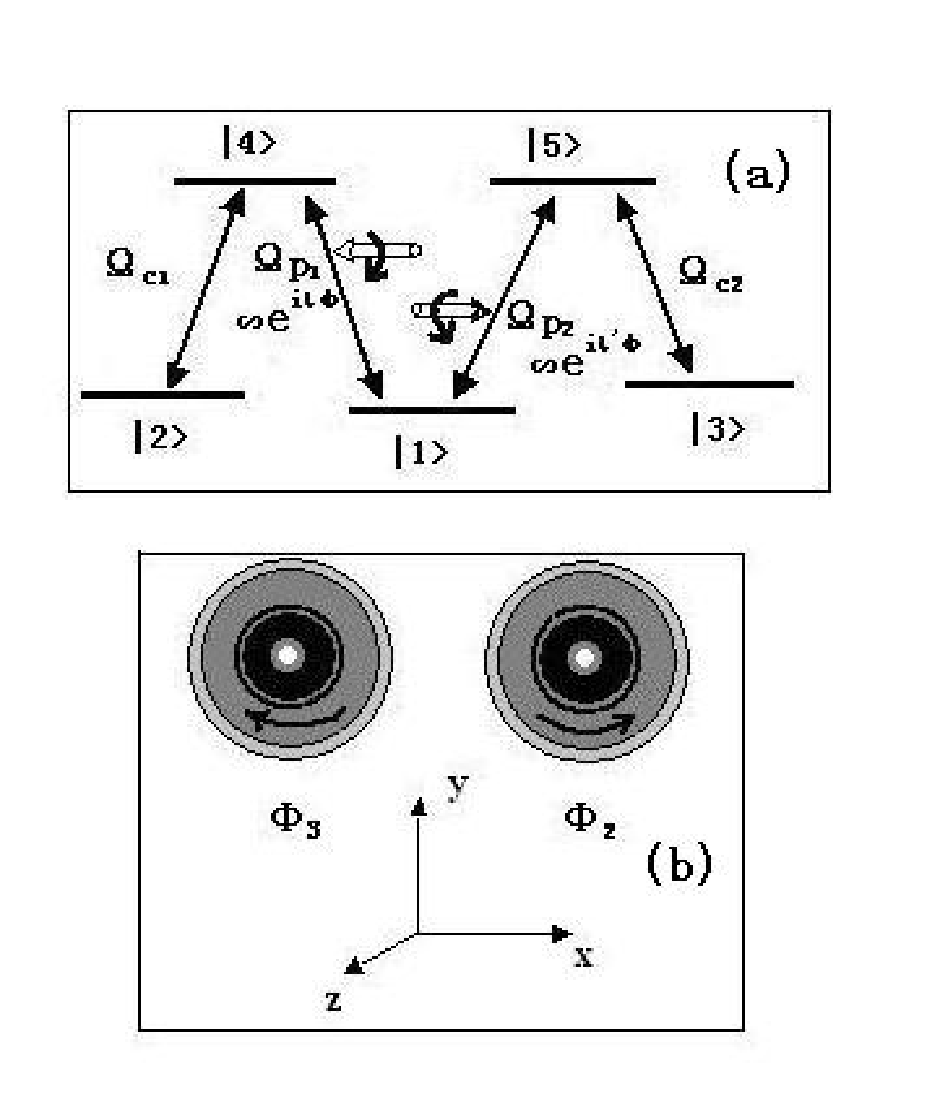}
\caption{(a)The condensate atoms with internal five-level $M$ type
configuration interact with two probe and two control beams, where
the two probe beams have orbital angular momentum $\hbar l$ and
$\hbar l'$, respectively in the $z$ direction. (b)Output
two-flavor atom lasers in corresponding vortex states (in $\pm z$
vortical direction).} \label{}
\end{figure}

\end{document}